**Optofluidic sensor system with Ge PIN photodetector for CMOS-compatible sensing**

L. Augel1#, F. Berkmann1, D. Latta2, I. A. Fischer1, S. Bechler1, Y. Elogail1, K. Kostecki1, K. Potje-Kamloth2 and J. Schulze1

1 Institute of Semiconductor Engineering, University of Stuttgart, Stuttgart, Germany

2 Fraunhofer ICT-IMM, Mainz, Germany

#e-mail: augel@iht.uni-stuttgart.de, phone: +49 711 685 69200

Abstract: Vertical optofluidic biosensors based on refractive index sensing promise highest sensitivities at smallest area foot-print. Nevertheless, when it comes to large scale fabrication and application of such sensors, cheap and robust platforms for sample preparation and supply are needed – not to mention the expected ease of use in application. We present an optofluidic sensor system using a cyclic olefin copolymer microfluidic chip as carrier and feeding supply for a complementary metal-oxide-semiconductor-compatibly fabricated Ge PIN photodetector. Whereas typically only passive components of a sensor are located within the microfluidic channel, here the active device is directly exposed to the fluid, enabling top-illumination. The capability for detecting different refractive indices was verified by different fluids with subsequent recording of the optical responsivity. All components excel in their capability to be transferred to large scale fabrication and further integration of microfluidic and sensing systems. The photodetector itself is intended to serve as a platform for further sophisticated collinear sensing approaches.

---

Optical biosensors based on refractive index sensing concepts utilize the sensitivity of structures such as waveguides (Blanco et al. 2006), microring resonators (De Vos et al. 2007), plasmonic nanoantennas (Chen et al. 2011), and plasmonic nanohole arrays (Brolo et al. 2004) to changes in the surrounding refractive index induced by analyte-ligand binding events (Damborský et al. 2016). The resulting change in optical response can be used for highly sensitive detection of biomolecules. Refractive index sensors support on-chip integration with signal conditioning circuits as well as wireless transmission for ultra-compact sensing solutions. Thus, setups that combine microrings for optical sensing with integrated Ge detectors have been realized experimentally (Zang et al. 2015). For even more compact integrated sensors, plasmonic nanohole arrays (Guyot et al. 2011), plasmonic gratings (Patskovsky and Meunier 2013; Perino et al. 2014) and nanodiscs (Mazzotta et al. 2010) have been positioned directly onto Si-based photodetectors for on-chip electrical detection of the optical response. Besides their ability to analyze minute sample volumes, such sensors offer additional advantages of small device footprints and, as a consequence, low fabrication cost. Their vertical illumination eliminates furthermore the need for optical coupling into waveguide structures.

While most of the experimental work is dedicated to the improvement of the optical sensing sensitivity, much less effort has been devoted to aspects of system integration (Laplatine et al. 2017). For practical on-chip applications, sensors consisting of plasmonic structures situated on top of photodetectors have to be placed directly into microfluidic channels; this contrasts with interferometer-based solutions, in which only passive components of the sensor are in contact with fluids. Complications that can arise from a placement of active photodetectors within the channel include leakage of the microfluidic channel due to surface structuring necessary in the device fabrication process as well as short-out of the device if electrolytes are used as fluidic media. Here, we investigate the viability of a microfluidic system consisting of a cyclic olefin copolymer (COC) microfluidic chip for application with on-chip Ge-based diodes as complementary metal-oxide-semiconductor (CMOS) compatible photodetectors. Cost and integration aspects are primary motivations for the choice of our materials. Figures of merit for photonic biosensors concern the limit of detection, multiplexibility and fabrication cost (Erickson et al. 2008). Sensor fabrication costs, in particular, can be reduced if the well-established CMOS technology can be used. CMOS-compatible sensor fabrication comprises furthermore the advantageous possibility of manufacturing the photonic biosensor with high yield and at a low price per unit as well as the possibility to integrate it simultaneously into read-out circuits and data processing units maximizing the integration level. Additionally, in large scale fabrication, joining as well as machining processes have to be taken into account since they are expensive and limiting the yield.

Beyond the implementation of CMOS fabrication technology, polymeric materials play a major role in the field of integrated sensors due to their broad variation in their properties and the wide range of prototype fabrication techniques like e.g. milling or hot embossing. Moreover, the availability of cheap mass fabrication techniques like injection-molding opens the field of disposable devices.

In the early development stage of integrated sensors, polydimethylsiloxane (PDMS) provides a rapid way to carry out preliminary tests enabled by its straightforward processability via casting as well as its self-sealing properties due to its softness. More rigid materials like COC or polycarbonate (PC) have the advantage, in comparison to PDMS, that the microfluidic channel cross-section or the overall shape is relatively pressure resistant and, hence, better suitable for systems, in which a hydraulic force is used for fluid transportation. Nevertheless the integration of rigid sensors is more complex, because a sealing material has to be additionally integrated to achieve a leak tight sensor packaging.

In this work COC was chosen as fluidic chip material due to its high transparency over a broad wavelength range that allows in future applications the light source to be placed outside of the microfluidic channel, which minimizes malfunctions of the illumination source. In addition, the low water absorption, the high compressive strength in comparison to PMDS, and the possibility to upscale the fabrication by using injection moulding are advantageous for our application.

Our optofluidic sensor system is shown schematically in Fig. 1. The microfluidic channels were realized by three components (see Fig. 1b), which are the microfluidic chip, the adhesive tape and the Si-substrate. Tubing connectors and feeding channels were milled into the COC microfluidic chip, which was supplied by TOPAS® and covered with a pressure-sensitive adhesive (Absolute QPCR Seal (AB-1170)) ordered from Thermo Scientific®. The channels conducting the fluid over the photodetectors were formed in double-sided adhesive tape by ablation with a $CO_2$ laser. Using the double-sided adhesive tape, the Si-substrates with a feed size of 10 mm were glued manually into a pocket milled into the microfluidic chip, taking care to align the channels in the adhesive tape with the channels in the microfluidic chip. The electrical contacts are still readily accessible from one side since the microfluidic chip does not fully cover the Si-substrate (Fig. 1a).

A schematic cross-section through the active device region of the Ge PIN photodetector is shown in Fig. 1e. The semiconductor layers are grown by a solid-source molecular beam epitaxy system on a $p^-$-Si substrate. Layer growth started with the deposition of a 100 nm B-doped Ge-layer that was annealed at 850 °C to form a virtual substrate in order to enable the subsequent growth of high-quality Ge-layers (Kasper et al. 1998). Layer growth continued with the deposition of a 300 nm p-doped Ge-layer and a 300 nm intrinsic Ge-layer followed by a top contact layer consisting of 100 nm n-doped Ge and 100 nm n-doped Si. The topmost layer is composed of Si in order to facilitate contacting the device. The doping concentrations in the p-doped region as well as in the n-doped region are $10^{20}$ cm$^{-3}$. A vertical layer growth approach has been used in several concepts that necessitate the integration of Ge-photodetectors on a Si-platform and is also available as foundry process in the context of group-IV photonic devices for communication technologies (Beals et al. 2008). In industry-scale wafer production, vertical layer growth is realized by differential epitaxy using a plasma-enhanced chemical vapor deposition (PECVD) process in the front end of line (FEOL) of fabrication.

Device fabrication started with mesa-structuring using standard photolithography and an inductively coupled plasma reactive-ion etching (ICP-RIE) process. The mesa height of only 600 nm can be expected to have negligible influence on the properties of the microfluidic channel. After passivation with $SiO_2$ from a PECVD-process and opening of contact holes defined through a lithography process, sputtered Al was used as a CMOS-compatible metal contact-layer and structured with photolithography and dry-etching. The samples were diced into 10x10 mm chips containing 2x10 diodes aligned for use in two microfluidic channels (see Fig. 1c and Fig. 1d).

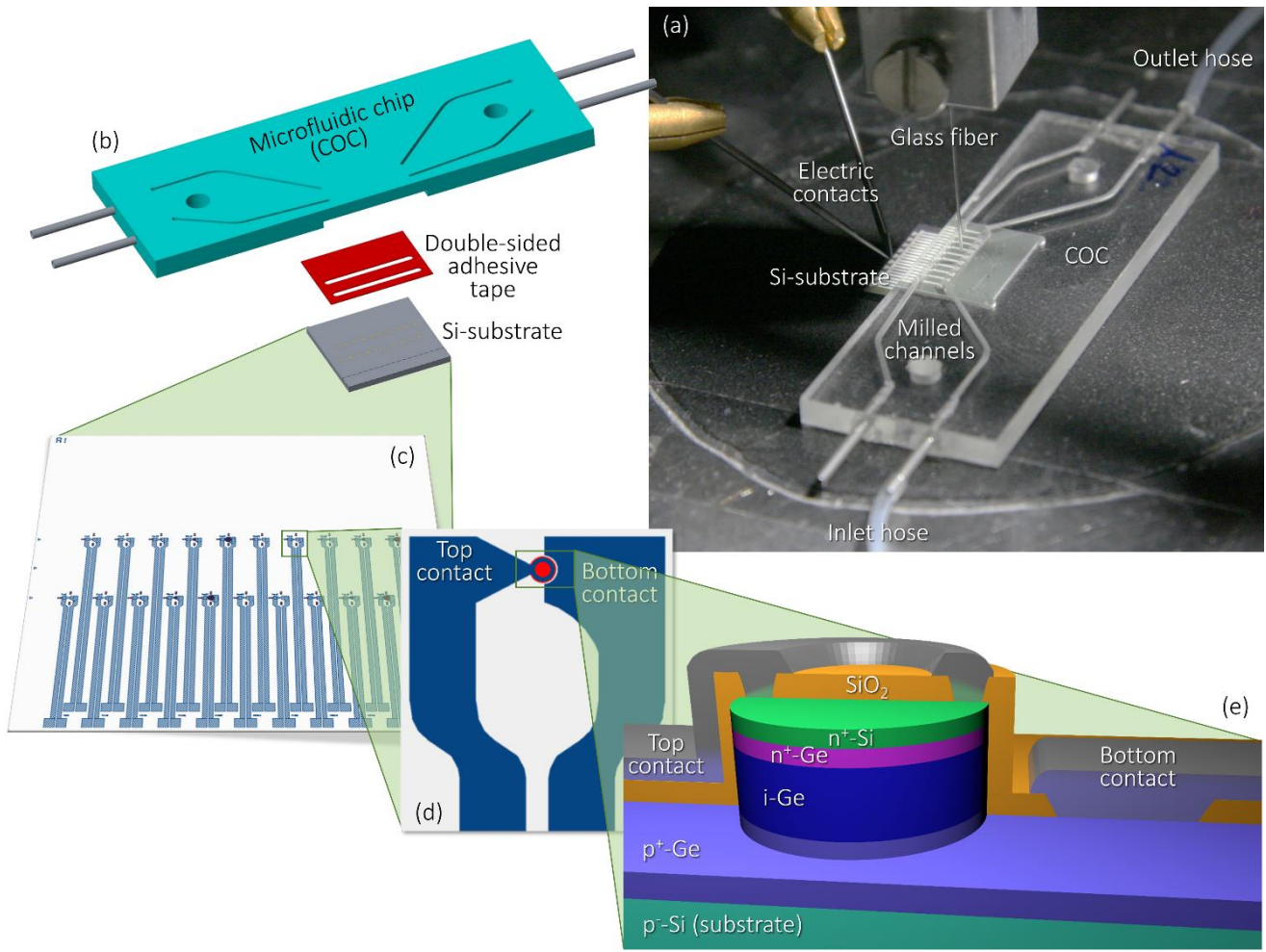

Fig. 1: Details of the optofluidic sensing system. (a) Image of the optofluidic sensing system under test. (b) Complete microfluidic setup with microfluidic chip with milled microfluidic channels and tubing connectors for inlet and outlet from the two microfluidic lines. Using the double-sided adhesive tape, with the two fluidic lines fabricated by laser ablation, the prepared Si-substrate is glued into a cavity inside the microfluidic chip. (c) Mask layout of the Si-substrate with 10 x 10 mm with two lines of photodetectors for use inside two separated microfluidic channels and contact metallization. (d) Mask layout of one single photodetector. (e) Schematic 3D plot of the Ge PIN photodetector showing the MBE-layers as well as passivation and Al-metallization.

The optofluidic sensing system was tested by filling the microfluidic channels sequentially with different fluids (Ethanol, Isopropanol and DI-water; all "very large scale integration" (VLSI) quality) and recording the optical response of the photodetector under illumination with light of different wavelengths. The illumination was provided by an external light source and coupled into the sensor via a glass fiber in vertical incidence (see Fig. 1a) whose alignment relative to the detector remained fixed for the complete duration of the measurements performed on the device. The glass fiber was fed by a super-continuum light-source equipped with an acousto-optical tunable filter able to supply narrow output spectra over a broad wavelength range. The photodetector was operated in photovoltaic mode (with external bias $V_{\text{bias}} = 0$) and diode currents with ($I_{\text{light}}(\lambda)$) and without ($I_{\text{dark}}(\lambda)$) illumination were measured with an on-wafer semiconductor analyzer for incident wavelengths in the range from 1250 nm to 1600 nm and in steps of 5 nm. The responsivity $R_{\text{opt}}(\lambda)$ of the devices is obtained by dividing the photocurrent $I_{\text{light}}(\lambda) - I_{\text{dark}}(\lambda)$ by the incident optical power $\Phi_{\text{opt}}(\lambda)$ pointwise:

$$R_{\text{opt}}(\lambda) = \frac{I_{\text{light}}(\lambda) - I_{\text{dark}}(\lambda)}{\Phi_{\text{opt}}(\lambda)}$$

The incident optical power $\Phi_{\text{opt}}(\lambda)$ is monitored by measuring it after each measurement of diode currents $I_{\text{light}}(\lambda)$ and without $I_{\text{dark}}(\lambda)$ at a fixed wavelength. The main source of experimental error in our setup is given by the measurement accuracy of the semiconductor analyzer ($\pm 0.1$ %) leading to error bars smaller than the symbol size in the plots shown in Fig. 2.

Since the radiation has to penetrate through the microfluidic chip as well as the sample volume before reaching the photodetector the recorded photocurrents differ depending on the refractive index of the fluid filling the microfluidic channel and absorption within the fluid.

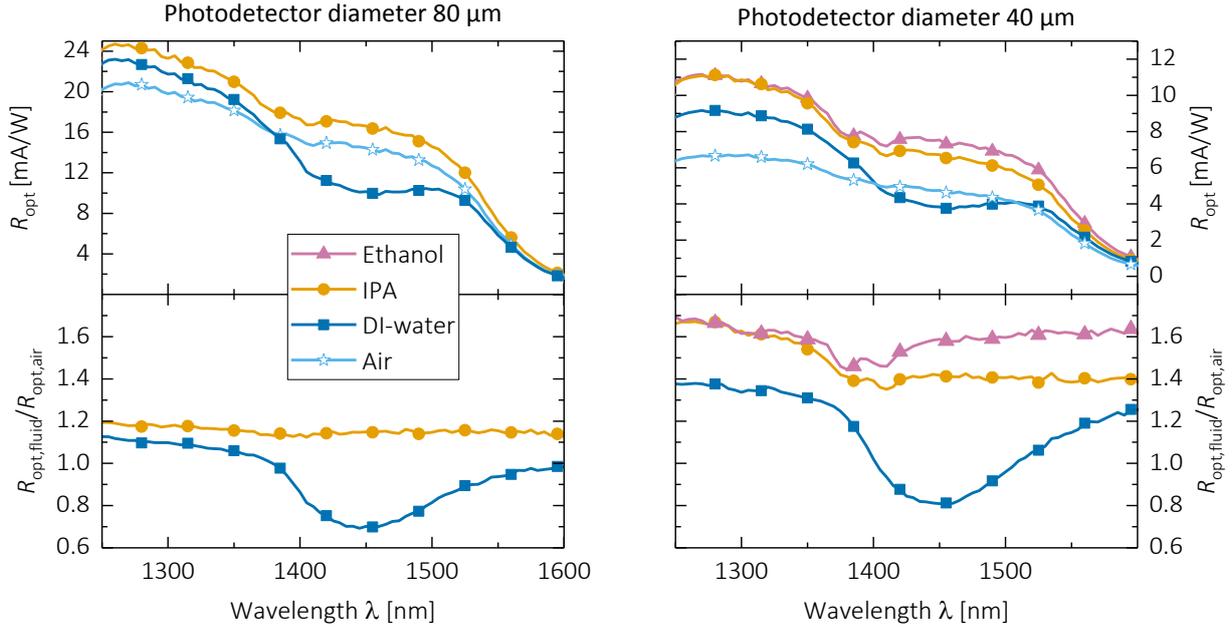

Fig. 2: Responsivities measured on a photodetector with 80 µm diameter and 40 µm diameter under different microfluidic conditions. Lower graphs show the responsivities of the fluids normalized to the responsivities of an air filled fluidic channel.

Figure 2 shows the responsivity for two different photodetectors with a diameter of 80 µm and 40 µm when filling the channel with the previously mentioned fluids or keeping it empty (air). It can be clearly seen that filling the channel with a fluid of higher refractive index increases the responsivity since the reflection at the COC-channel as well as channel-semiconductor interface decreases compared to an air-filled channel. Furthermore, as shown in Fig. 2 for the photodetector with a diameter of 40 µm, the change in responsivity when introducing either ethanol (with refractive index $n_{EtOH}$ = 1.353) or isopropanol ($n_{IPA}$ = 1.375) into the channel is large compared to an air-filled channel ($n_{air}$ = 1). By comparison, the responsivities obtained for an isopropanol-filled channel and for an ethanol-filled channel are almost identical for wavelengths below 1400 nm as a result of the small difference in refractive indices of the two fluids. Finally, the characteristic absorption of water ($n_{DI}$ = 1.321) around 1450 nm lowers the responsivity in this wavelength region, making it clearly distinguishable from the other fluids. Additionally, a feature from the microfluidic chip is visible at 1400 nm where COC absorbs in a small band (compare to Fig. 3 (TOPAS Advanced Polymers 2017)).

When normalizing the responsivity obtained from a filled microfluidic channel to an empty channel's responsivity, i.e. when calculating

$$V = \frac{R_{\text{opt,fluid}}}{R_{\text{opt,air}}}$$

the different fluids can clearly be differentiated (lower graphs in Fig. 2). The higher ratio observed at the photodetector with a diameter of 40 µm can be attributed to the lower refraction at the above mentioned COC-channel interface when filling the channel with a fluid, funneling more light into the detector. This aspect can also be recognized from the lower responsivity measured at this detector since the initially collimated incident light is scattered at the air-COC interface, widening the beam cone such that it becomes larger than detector surface. We note that for this reason, glass fiber alignment with respect to the detector has to be kept fixed during measurements in order to be able to compare responsivity spectra for different fluids present in the microfluidic channel.

This proof-of-concept of a simplistic refractive index sensor can be used as a platform for further improved CMOS-devices, which are e.g. equipped with localized surface plasmon resonance (LSPR) based sensing approaches using metallic nanostructures (Canalejas-Tejero et al. 2014; Augel et al. 2016; Fischer et al. 2016). The platform even offers the feasibility of the complex flow-through instead of the classical flow-over LSPR sensing unit approach (Escobedo et al. 2010). Here, shorter sensor response times can be achieved, by circumventing the mainly convective mass transport to the sensing surface in flow-over sensors. Combining the refractive index detector and transducer within the same device would further decrease sample preparation complexity and overall size (Cetin et al. 2014) of such systems.

In this work, we found that it is possible to use a microfluidic chip made from COC as a carrier for a Ge PIN photodetector fabricated in CMOS compatible technology. In our setup, the detector for a vertical and top-illuminated optofluidic sensing unit is completely located within the microfluidic channel. Utilizing this setup for future integrated biosensors can be expected to result in a reduction of the price per unit since production costs in CMOS fabrication depend on the sensor area and large-scale fabrication of the microfluidic chip is possible by injection molding. Absorption within the COC material used for the microfluidic waveguides does not limit the sensing capabilities: our simple sensor was able to distinguish between fluids with different refractive indices flowing over the detector. Within the scope of future work the detector will be equipped with a plasmonic nanohole array combining high-sensitivity sensing using LSPR (Augel et al. 2016) and detection within one device.

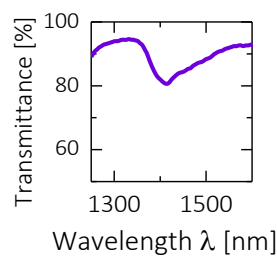

Fig. 3: Transmittance through a 2 mm slab of COC.

Acknowledgement: This work was supported through funding by the University of Stuttgart and the Ministry of Science and Education of Baden-Wuerttemberg (RiSC). Furthermore, we appreciate the support by the Institute of Electron Devices and Circuits at Ulm University especially from S. Jenisch and S. Strehle.